\documentclass[letterpaper,twocolumn,10pt]{article} 

\usepackage{zhanggroup}
\usepackage{hyperref}       
\usepackage{url}            
\usepackage{booktabs}       
\usepackage{amsfonts}     
\usepackage{pifont}
\usepackage{nicefrac}       
\usepackage{microtype}     
\usepackage{tikz}
\usepackage{xspace} 
\usepackage{amsmath}
\usepackage{mathrsfs}
\usepackage{amssymb}
\usepackage{amsthm}
\usepackage{subcaption}
\usepackage{multirow}
\usepackage{makecell}
\usepackage{wrapfig}
\usepackage{lipsum}

\newcommand{\m}{\mathcal{M}}
\newcommand{\mt}{\mathcal{M}_{\textit{target}}}

\newcommand{\D}{\mathcal{D}}

\newcommand{\dlnm}{\mathcal{D}^{\textit{non\_member}}_{\textit{local}}}

\newcommand{\da}{\mathcal{D}_{\textit{auxiliary}}}

\newcommand{\dtm}{\mathcal{D^{\textit{member}}_{\textit{sub\_target}}}}

\newcommand{\iq}{\boldsymbol{x}}
\newcommand{\ig}{\boldsymbol{x'}}
\newcommand{\tg}{\boldsymbol{t}}
\newcommand{\mypara}[1]{\noindent{\bf {#1}.}}

\title{Membership Inference Attacks Against Text-to-image Generation Models}

\author{
Yixin Wu\textsuperscript{1}\ \ \
Ning Yu\textsuperscript{2}\ \ \
Zheng Li\textsuperscript{1}\ \ \
Michael Backes\textsuperscript{1}\ \ \
Yang Zhang\textsuperscript{1}
\\
\\
\textsuperscript{1}\textit{CISPA Helmholtz Center for Information Security}\ \ \ 
\textsuperscript{2}\textit{Salesforce Research}\ \ \
}
\date{}

\begin{document}

\maketitle

\begin{abstract}
Text-to-image generation models have recently attracted unprecedented attention as they unlatch imaginative applications in all areas of life.
However, developing such models requires huge amounts of data that might contain privacy-sensitive information, e.g., face identity.
While privacy risks have been extensively demonstrated in the image classification and GAN generation domains, privacy risks in the text-to-image generation domain are largely unexplored.
In this paper, we perform the first privacy analysis of text-to-image generation models through the lens of membership inference. 
Specifically, we propose three key intuitions about membership information and design four attack methodologies accordingly.
We conduct comprehensive evaluations on two mainstream text-to-image generation models including sequence-to-sequence modeling and diffusion-based modeling.
The empirical results show that all of the proposed attacks can achieve significant performance, in some cases even close to an accuracy of 1, and thus the corresponding risk is much more severe than that shown by existing membership inference attacks.
We further conduct an extensive ablation study to analyze the factors that may affect the attack performance, which can guide developers and researchers to be alert to vulnerabilities in text-to-image generation models.
All these findings indicate that our proposed attacks pose a realistic privacy threat to the text-to-image generation models.
\end{abstract}

\section{Introduction}

With the superb power of unfastening limitless imaginative content creations, text-to-image generation has become one of the most noteworthy topics in the computer vision field and has been advanced significantly in recent years by a series of designs such as sequence-to-sequence based models (e.g., Parti~\cite{YXKLBWVKYAHHPLZBW22}), and diffusion-based models (e.g., DALL-E 2~\cite{RDNCC22} and Imagen~\cite{SCSLWDGAMLSHFN22}). 
Along with the extremely rapid development of this topic, the demand for data is highly growing.
For instance, the Parti/DALL-E 2/Imagen models are trained on 6.6B/650M/860M image-text pairs, respectively.
A stark reality is that such large amounts of training data collected for model builders often contain inherently privacy-sensitive information, such as facial identity, and have raised community concerns.
Under the terms of the GDPR,\footnote{\url{https://gdpr-info.eu/}} the LAION organization is calling on people to determine whether their private information exists in publicly released LAION datasets and providing support for the removal of their private data.\footnote{\url{https://laion.ai/gdpr/}}
Unfortunately, model builders are unlikely to disclose their training data due to the huge effort and resources they have put into collecting them~\cite{UNSS17, RCK18, ZGJWSHM18,ABCPK18,LHZG19}. 

Various recent studies have shown that machine learning (ML) models are vulnerable to privacy attacks against their training data, and a major attack in this area is membership inference: an adversary aims to infer whether a data sample is part of the dataset trained by the target ML model.
The resulting privacy leakage caused by this attack would raise serious issues as the training data is intellectual property and contains sensitive information. 
In addition, data owners can also use it to audit whether their data was collected by model builders without authorization under GDPR terms, i.e., having better control over their data.

Existing membership inference attacks have been demonstrated to be a realistic threat to different type of tasks, such as classification~\cite{SSSS17, SZHBFB19, HYYBGC21, CTCP21, LZ21,ONK21, HWWBSZ21, WYPY21, LLHYBZ222,HLXCZ22} and GAN generation~\cite{HHB19, CYZF20}.
Unfortunately, the peculiarities of text-to-image generation do not allow to trivially extend the understanding of membership leakage from fully explored classification and GAN generation domains to text-to-image domain.
Hence, these aforementioned realities motivate us to focus on membership leakage in text-to-image generation models.

\mypara{Contribution}
In this work, we take the first step towards studying membership leakage in text-to-image generation models, where an adversary aims to infer whether a given image is used to train a target text-to-image generation model.
In particular, we focus on the most difficult and realistic scenario where no additional information about the target model is available to the adversary other than the output images.
Based on the characteristics of the text-to-image generation models, we consider three key intuitions and design four attack methods accordingly.
We conduct comprehensive experiments on two representative text-to-image generation modes, i.e., sequence-to-sequence and diffusion-based.
Extensive empirical results show that all of our proposed attack methodologies achieve remarkable performance, which convincingly demonstrates that membership leakage is a severe threat to the text-to-image generation models.
Furthermore, to delve into which factors and their impact on the attack performance, we conduct a comprehensive ablation study from different perspectives that can guide developers and researchers to be alert to vulnerabilities in text-to-image generation models.
Our main contributions are as follows:
\begin{itemize}
    \item We pioneer in studying the privacy risks of text-to-image generation models from the perspective of membership inference.
    \item We consider three intuitions and design four attack methodologies via differentiating attack intuitions. 
    \item We conduct an extensive evaluation on two mainstream text-to-image generation models, and the results show the effectiveness and generalizability of the proposed attacks, indicating such membership leakage poses a much more severe threat than existing work.
    We further perform a comprehensive ablation study from different angles to analyze the factors that may affect the attack performance, which is expected to provide instructional warnings to model inventors.
\end{itemize}

\section{Background}
\label{section:background}

\subsection{Text-to-image Generation Models}

Text-to-image generation targets to unlock the innovative applications covering various areas of life, including painting, design, and multimedia content creation.  
It generates creative images combining concepts, attributes, and styles, from expressive text descriptions. 
Currently, text-to-image generation models can be divided into two different designs, namely diffusion-based modeling and sequence-to-sequence modeling.

\mypara{Diffusion-based}
The diffusion-based models directly leverage noise as the input of the de-noising network, starting from these random points and gradually de-noising them conditioned on textual descriptions until images matching the conditional information are generated.
Building on the power of diffusion models in high-fidelity image synthesis, the text-to-image generation is significantly pushed forward by the recent effort of GLIDE~\cite{NDRSMMSC21}, LDM~\cite{RBLEO22}, DALL-E 2~\cite{RDNCC22} and Imagen~\cite{SCSLWDGAMLSHFN22}.

\mypara{Sequence-to-sequence} 
The main idea of this design is to turn images into discrete image tokens via leveraging transformer-based image tokenizers (e.g., dVAE~\cite{R17}), and employ the sequence-to-sequence architectures to learn the relationship between textual input and visual output from a large collection of text-image pairs. 
The representative works of sequence-to-sequence modeling are Parti~\cite{YXKLBWVKYAHHPLZBW22}, DALL-E~\cite{RPGGVRCS21} and CogView~\cite{DYHZZYLZSYT21}.

In this work, we adopt LDM and DALL-E mini as our target models, representing diffusion-based and sequence-to-sequence models, respectively.

\subsection{Membership Inference Attacks}

Membership inference attacks (MIAs), aiming to infer whether a specific data sample was involved in a target model's training phase (called member or non-member), are considered as an approach to investigating privacy leakage and detecting illegal data abuse.
The basic idea is to exploit the behavioral difference of the target model on members and non-members.
Depending on the characteristics of the target model, the behavioral differences can be constructed in different ways.
For instance, in the classification domain, the behavioral difference exploited in most prior works~\cite{SSSS17, SZHBFB19, HYYBGC21,ONK21, HWWBSZ21, WYPY21} is the confidence score of members over non-members.
More recent works~\cite{LZ21, CTCP21} later attack in a more realistic scenario where the adversary has access only to the predicted labels of the target model.
Here, the behavioral differences refer to the fact that the perturbations added to members to change the predicted labels are larger than those of non-members.
In image generation domain where the models accept random latent code as input and then output images, Chen et al.\  and Hilprecht et al.\ propose a customized attack method that estimates the probability of the query sample can be generated by the generator, where the behavioral difference is that the probability of members is greater than non-members~\cite{CYZF20, HHB19}.

Unfortunately, these popular and well-explored attack methodologies in classification and image generation domains cannot be trivially extended to the text-to-image generation domain, because the text-to-image generation model accepts text as input and then outputs images, which is totally different from previous works.
Hence, it is difficult for existing attack methods to evaluate whether text-to-image generation models are truly vulnerable to membership inference, which prompts the need to investigate new attack methods specifically for the text-to-image generation models.

\begin{figure*}[!t]
\centering
\includegraphics[width=2\columnwidth]{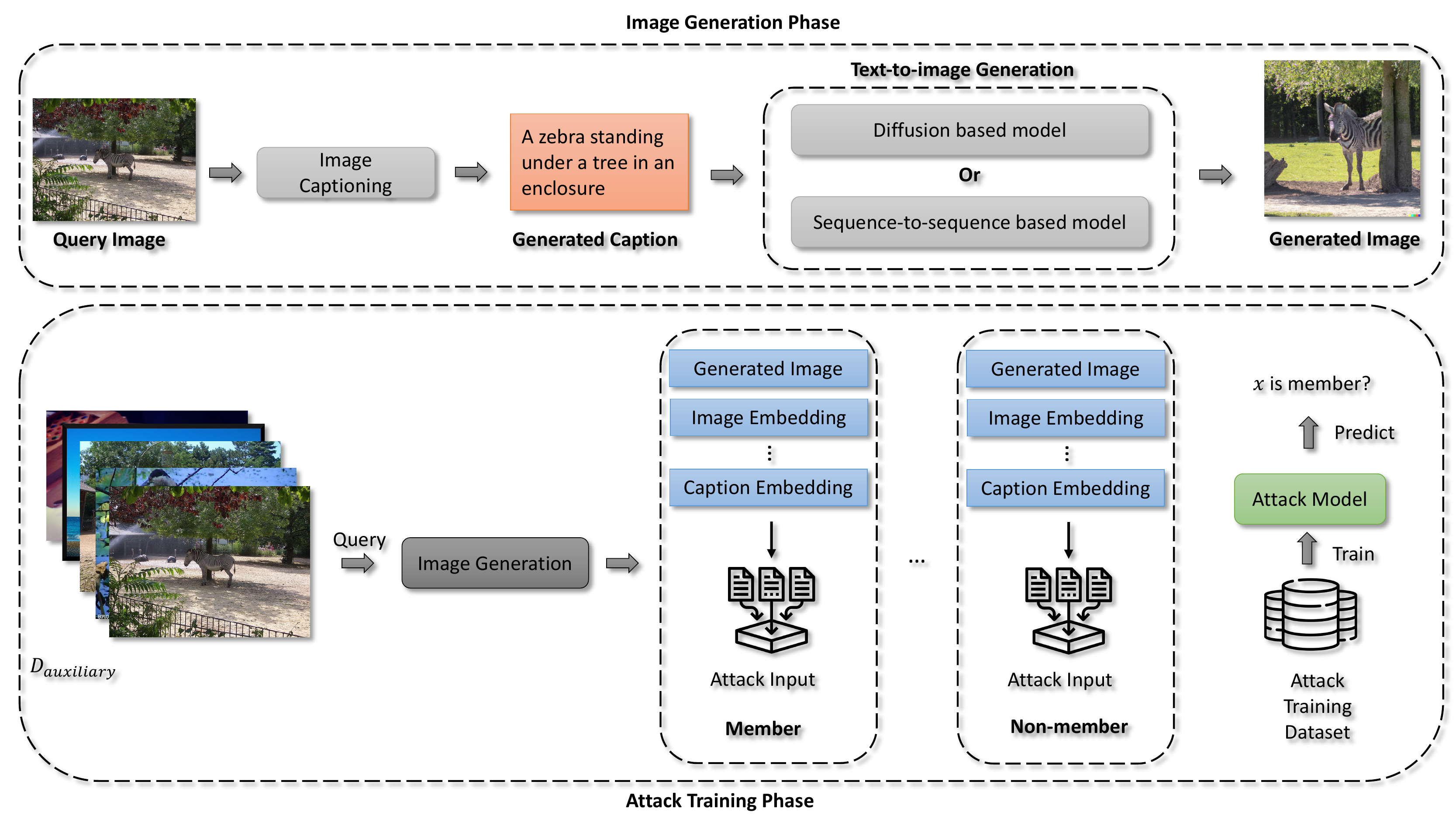}
\caption{Overview of our attack pipeline.}
\label{figure:pipeline}
\end{figure*} 

\section{Problem Statement}

In this section, we formulate the text-to-image generation and the threat model.

\subsection{Text-to-image Generation}

As aforementioned, we focus on the membership leakage in the text-to-image generation domain.
A text-to-image generation model $\m$ can map a text caption $t$ to the corresponding image $\iq$.
To construct a text-to-image generation model $\m$, one needs to collect a huge amount of data pairs ($\boldsymbol{t}$,  $\iq$) to construct the training set $\D$.
The model is then optimized via minimizing a predefined loss function.

\subsection{Threat Model}

\mypara{Adversary's goal}
The goal of the adversary is to infer whether the user's image $\iq$ is used to train a target text-to-image generation model $\mt$.

\mypara{Adversary’s knowledge}
Typically, the training datasets are composed of a huge number of data pairs, i.e., text caption $\tg$ and the corresponding image $\iq$.
Here, we assume that the adversary only queries a candidate image $\iq$ without its corresponding text caption to infer the membership, which is more realistic and broadly applicable.
We assume the adversary only has black-box access to the target model $\mt$, which is the most difficult and realistic scenario.
Besides, we assume that the adversary has a very small subset from the member training data of target model $\dtm$, as well as a small set of local non-member data $\dlnm$.
The adversary then constructs an auxiliary dataset $\da = \{x^m \cup x^{nm}: x^m \in \dtm, x^{nm} \in \dlnm \} $ that can be used to train the attack model $\mathcal{A}$, i.e., a binary classifier.
Note that, the assumption of auxiliary dataset also holds for previous works~\cite{LWHSZBCFZ22}.
The reason of the second assumption is that due to the resource limitation, i.e., insufficient GPU resource, so we are unable to leverage shadow technique~\cite{LLHYBZ222, LZBZ22}, i.e., training shadow models on billions of local image-caption pairs to mimic the target model $\mt$. 

\section{Methodology}
\label{section:method}

In this section, we start with the design intuition, and then we introduce the attack methodologies.

\subsection{Intuitions}

The key intuition of our work is a general observation about the overfitting nature of ML models.
Concretely, given a query data pair (text caption $\tg$ and the corresponding image $\iq$), the text-to-image generation models accept text $\tg$ as input and are optimized to generate image $\ig$ that are similar or same to the original image $\iq$.
This derives to the following three perspective of key intuitions on membership information:

\mypara{Intuition I} The quality of generated image $\ig$ of data pair ($\tg$ and $\iq$)  from training set should be higher than that from testing set.

\mypara{Intuition II} The reconstruction error between generated image $\ig$ and original image $\iq$ from training set should be smaller than that from testing set.

\mypara{Intuition III} The generated image $\ig$ should more faithfully reflects the semantic of a textual caption $\tg$ from the training set than the testing set.

Therefore, we focus on the distinguishability of membership by observing behavioral differences of these intuitions, i.e., members and non-members behave differently in the three aspects mentioned above.

\begin{table*}[!t]
\centering
\caption{
Attack Taxonomy.
``\checkmark'' means this attack is based on the intuition and ``-'' indicates the intuition is not necessary.
}
\begin{tabular}{c|c|c|c|c|c} 
\toprule
\multirow{2}{*}{Attack Category}  & \multicolumn{2}{c|}{Intuition I} &\multicolumn{2}{c|}{Intuition II}& Intuition III \\
 & Pixel-level &  Semantic-level &Pixel-level &Semantic-level & Semantic-level \\
\midrule
Attack I-P & \checkmark & - & - & - & - \\
\cmidrule(lr){2-6}
Attack I-S & - & \checkmark & - & - & - \\
\cmidrule(lr){1-6}
Attack II-P   & - &  - & \checkmark & - & - \\
\cmidrule(lr){2-6}
Attack II-S   & - &  - & - & \checkmark & - \\
\cmidrule(lr){1-6}
Attack III   & - &  - & - & - & \checkmark \\
\cmidrule(lr){1-6}
Attack IV & - &  \checkmark & - & \checkmark &\checkmark \\
\bottomrule
\end{tabular}
\label{tab:attackoverview}
\end{table*}

\subsection{Attack Methodologies}

As the adversary only holds a query image $\iq$ that they aim to infer, the adversary initializes the attack by leveraging a third parity image captioning tool to generate a caption $\tg$ for the given query image $\iq$, as illustrated in~\autoref{figure:pipeline}.
Then, they feed the generated caption $\tg$ into the target text-to-image generation model $\mt$ to obtain a generated image $\ig$.
In this way, we connect the query image and generated images explicitly.
In addition, we only need to query the target model once for each query image to get one corresponding generated image, largely decreasing the possibility of being detected by defense mechanisms. 
Building on the attack pipeline, we design 4 different types of attacks via exploiting different intuitions, i.e., Attack-I/II/III based on Intuition-I/II/III, and Attack-IV using all intuitions, as illustrated by~\autoref{tab:attackoverview}.
In the end, the adversary constructs an attack dataset and trains the attack models.
The dataset is split by half as the attack training dataset and attack testing dataset.
We provide the details of each attack method as follows.

\mypara{Attack I} 
This attack is based on \textit{Intuition-I} that there is a discrepancy between members and non-members in terms of the quality of generated images.
For simplicity, an adversary can differentiate between members or non-members by feeding the query image directly into the attack model, rather than measuring quality and then making the distinction.
Such an attack method is based on the pixel-level discrepancy, hence is named \textit{Attack I-P}.
Besides the pixel-level discrepancy, we further consider a semantic-level discrepancy of generated images.
Concretely, the adversary applies a pre-trained vision-language models (e.g., CLIP~\cite{RKHRGASAMCKS21}) to extract the embedding of the generated images and then feeds the embedding into the attack model, which is referred to as \textit{Attack I-S}.

\mypara{Attack II}
This attack is based on the \textit{Intuition-II} that there is a discrepancy between members and non-members in terms of reconstruction errors.
In particular, we again consider both pixel-level and semantic-level reconstruction errors.
The former, \textit{Attack II-S}, directly measures the distance between the generated image and its corresponding query image, while the latter, \textit{Attack II-S}, also applies a pre-trained visual language model to extract the embeddings of the generated image and its corresponding query image, respectively, and then measures the distance between them.
Finally, the adversary feeds the distances into the attack model to distinguish between members and non-members.

\mypara{Attack III}
This attack is based on the \textit{Intuition-III} that there is a discrepancy between members and non-members in terms of faithful reflection.
Here, faithful reflection means that the generated images faithfully reflect the semantics of the text captions.
Therefore, we only consider the semantic level attack pipeline, i.e., the adversary first applies a pre-trained visual language model to extract the embedding of the generated images and their corresponding text captions separately, and then measures the distance between them.
Finally, the attacker feeds the distances into the attack model to distinguish between members and non-members.

\mypara{Attack IV}
This attack is based on all three intuitions.
For \textit{Intuition-I} and \textit{Intuition-II}, we only consider semantic-level discrepancies because our experiments on \textit{Attack-I} and \textit{Attack-II} demonstrate that semantic-level discrepancies perform much better than pixel-level discrepancies (see~\autoref{section:results}).
Thus, we take all attack features based on three semantic-level discrepancies of \textit{Intuition-I/II/III} as input and feed them into the attack model to distinguish between members and non-members.

\section{Evaluation}
\label{section:eval}

\subsection{Experimental Setup}

\mypara{Target models and training datasets}
We demonstrate the efficacy and generalizability of the proposed membership inference attack against two mainstream text-to-image generation models: LDM represented for diffusion-based modeling and DALL-E mini represented for sequence-to-sequence modeling.
We download pre-trained LDM\footnote{\url{https://github.com/CompVis/latent-diffusion}} and DALL-E mini\footnote{\url{https://github.com/borisdayma/dalle-mini}} models from the Web and evaluate our proposed attack methods directly on these real-world models, which is actually more realistic than almost all existing works only in a laboratory setting.

Each generation application benchmarks its own dataset.
LDM is pretrained on the LAION-400M dataset~\cite{SVBKMKCJK21}, DALL-E mini is pretrained on the CC3M~\cite{SDGS18}, CC12M~\cite{CSDS21}, and a filtered subset of YFCC100M~\cite{TSFENPBL16}.
All of them are widely used image-text pair datasets.

\begin{figure*}[!t]
\centering
\begin{subfigure}{0.9\columnwidth}
\includegraphics[width=0.9\columnwidth]{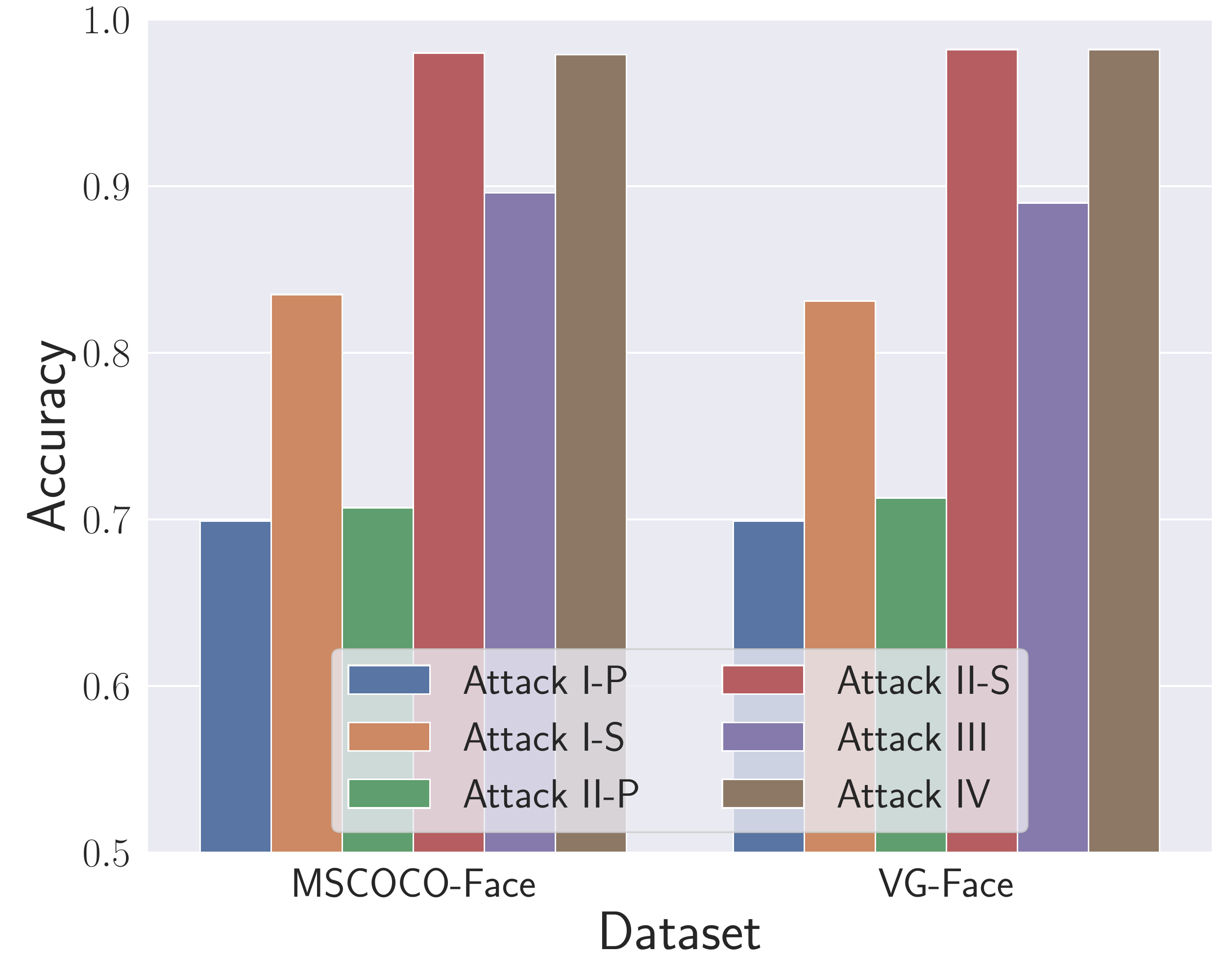}
\caption{LDM}
\label{figure:acc_ldm}
\end{subfigure}
\begin{subfigure}{0.9\columnwidth}
\includegraphics[width=0.9\columnwidth]{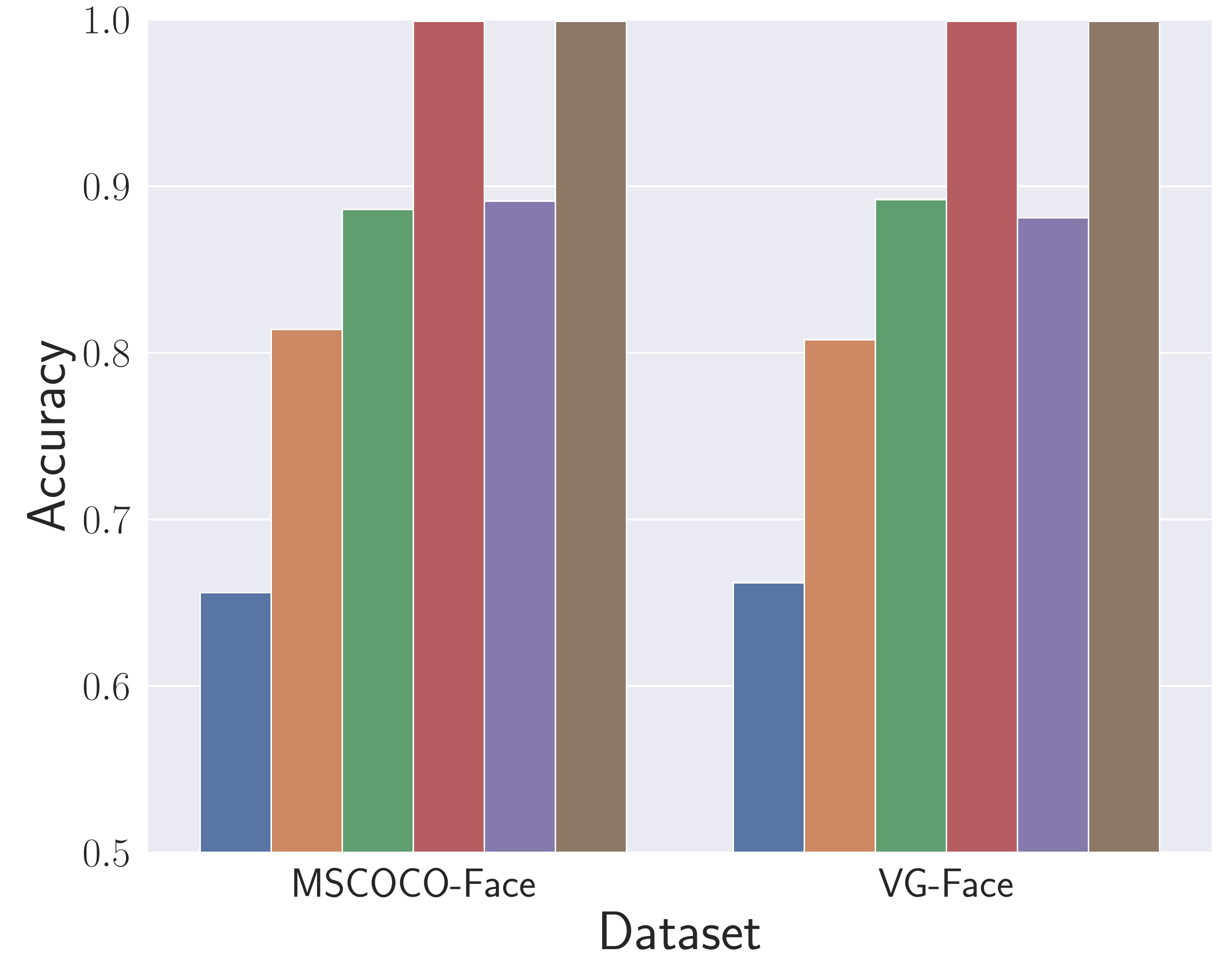}
\caption{DALL-E mini}
\label{figure:acc_dalle}
\end{subfigure}
\caption{Test accuracy of the proposed attack methods on the (a) LDM model and (b) DALL-E mini.
The caption and embedding generation tools for both cases are BLIP.}
\label{figure:accuracy_report}
\end{figure*}

\begin{table*}[!t]
\caption{FID scores between query images and its corresponding generated images for member and non-member datasets.}
\label{table:fid_score_analysis}
\centering
\renewcommand{\arraystretch}{1.2}
\centering
\begin{tabular}{c c c c}
\toprule
Member Dataset & FID Score &Non-member dataset & FID Score \\
\midrule
Laion-Face (30K) & 9.912 & MSCOCO-Face (30K) & 19.308 \\
Laion-Face (26K) & 9.959 & VG-Face (26K) & 20.314 \\
\bottomrule
\end{tabular}
\end{table*}

\mypara{Attack models and auxiliary datasets}
For pixel-level attacks, i.e., Attack I-P/II-P, the attack model is a CNN model.
For semantic-level attacks, i.e., Attack I-S/II-S/III, the attack model is a 3-layer MLP model.
The attack model of Attack-IV is composed of three sub-networks and a linear layer, as it has three types of input.
Every sub-network accepts one type of input and generates a feature embedding.
These three embeddings are concatenated and fed into a linear layer to make predictions.
The loss function is cross-entropy and the optimizer is Adam.
The learning rate is set to 0.001, and we train the attack model for 200 epochs.

To construct the auxiliary dataset, we randomly select a small set of member images from the target training set, as well as the same size non-member samples from two local datasets, i.e., MSCOCO~\cite{CFLVGDZ15} and Visual Genome (abbreviated as VG)~\cite{KZGJHKCKLSBF17}.
For the LDM model, with the intention of better focusing on the privacy-sensitive information, we leverage LAION-Face~\cite{ZYZBCHYCZW22}, a human face subset of LAION-400M, as members, and create MSCOCO-Face and VG-Face as non-members.
For the DALL-E mini model, as it is claimed that faces in general are not generated properly by the DALL-E mini model, we create CC3M-No-Face as members, MSCOCO-No-Face and VG-No-Face as non-members.
See more details about the dataset creation in~\autoref{section:appendix}.  
Then, we merge the member dataset and non-member dataset as the attack dataset and split it by half creating the attack training dataset and the attack testing dataset.

\mypara{Captions and embeddings generation tools}
We leverage officially pretrained BLIP~\cite{LLXH22}, a recent state-of-the-art vision-language model to generate captions, image embeddings, and text embeddings. 
We later show that the choice of the caption generation tool and embedding generation tool has negligible effect on the attack performance.

\mypara{Evaluation metrics}
We adopt accuracy as the main evaluation metric for the attack performance, as widely used in previous works~\cite{SSSS17, SZHBFB19}.

\subsection{Results}
\label{section:results}

\mypara{Attack performance} 
We first show the attack performance of our proposed attacks when treats MSCOCO and VG as the non-member datasets on LDM and DALL-E mini models in~\autoref{figure:accuracy_report}.
We can observe that all of the proposed attacks can achieve remarkable performance, even in the worst case with an accuracy of over 0.65, which is much higher than random guesses (i.e., 0.5).
These empirical results verify our key intuitions in all three perspectives.

Furthermore, we can also find that semantic-level attacks achieve much better performance than pixel-level attacks.
Take the LDM model as an example, the accuracy of Attack I-S/II-S on MSCOCO-Face is 0.835/0.980, while the accuracy of Attack I-P/II-P is 0.699/0.707.
Therefore, an adversary can apply semantic-level attacks to obtain higher performance if they have a semantic extraction tool, such as a well-trained BLIP.
We reason that this is due to the fact that these pixel-level features make pixels independent and are unable to obtain context-dependent information, whereas semantic-level features have this capability.
In addition, a query image can generate multiple images that are not identical at the pixel level but are similar at the semantic level, leading to significant variance in the performance of pixel-level attacks.
Meanwhile, the results of Attack II-S are much better than those of Attack III, indicating that the relationship between the same-modality embeddings works better than that of cross-modality embeddings.
These results lead to the unsurprising observation that the performance of Attack-IV and Attack II-S is very similar and that Attack IV and Attack II-S achieve the best attack performance in all cases.
Take the DALL-E mini model as an example, these two attacks even successfully achieve 99.9\% test accuracy on both non-member datasets.

\begin{figure*}[!t]
\centering
\begin{subfigure}{0.9\columnwidth}
\centering
\includegraphics[width=0.9\columnwidth]{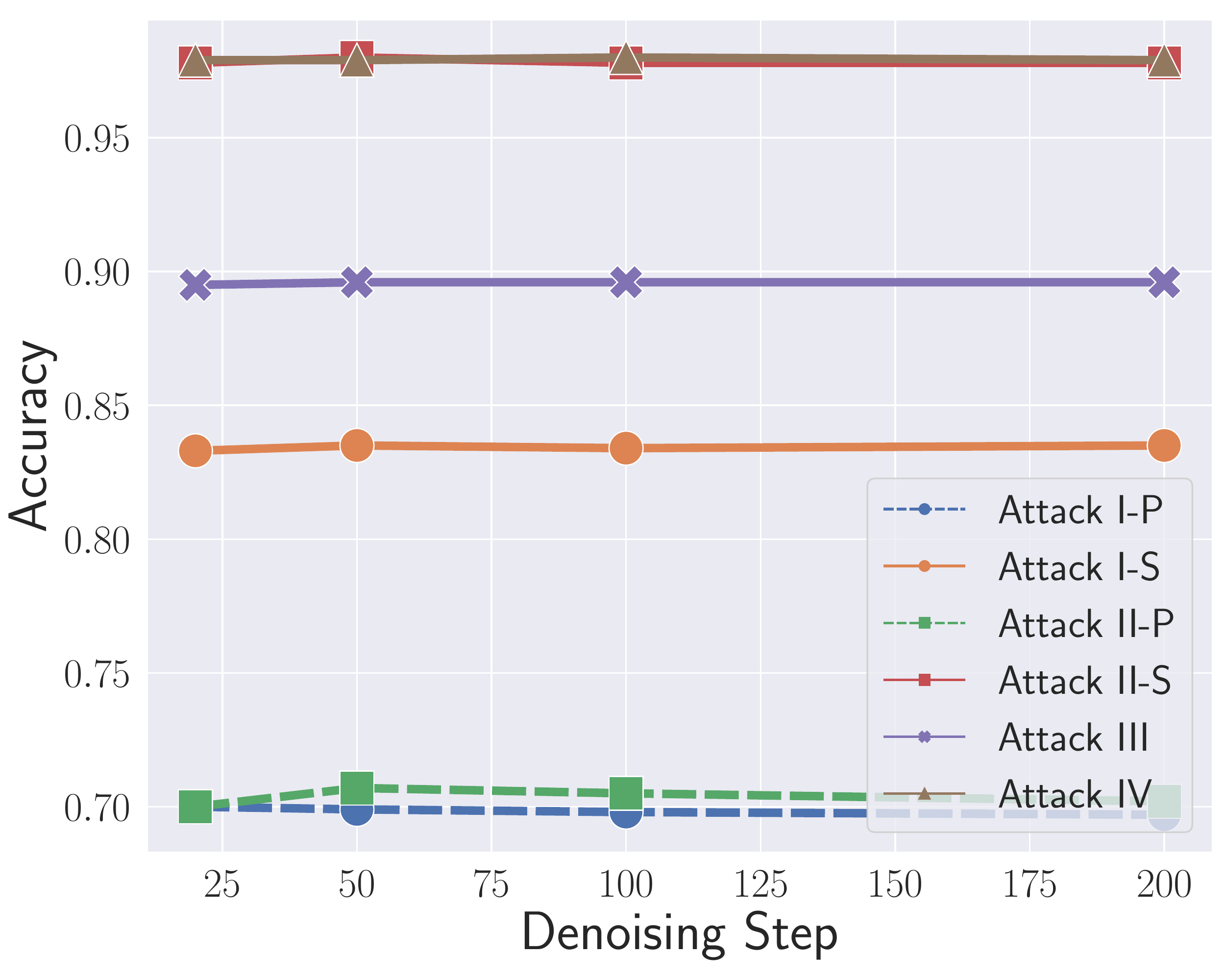}
\caption{MSCOCO-Face}
\label{figure:step_mscoco}
\end{subfigure}
\begin{subfigure}{0.9\columnwidth}
\centering
\includegraphics[width=0.9\columnwidth]{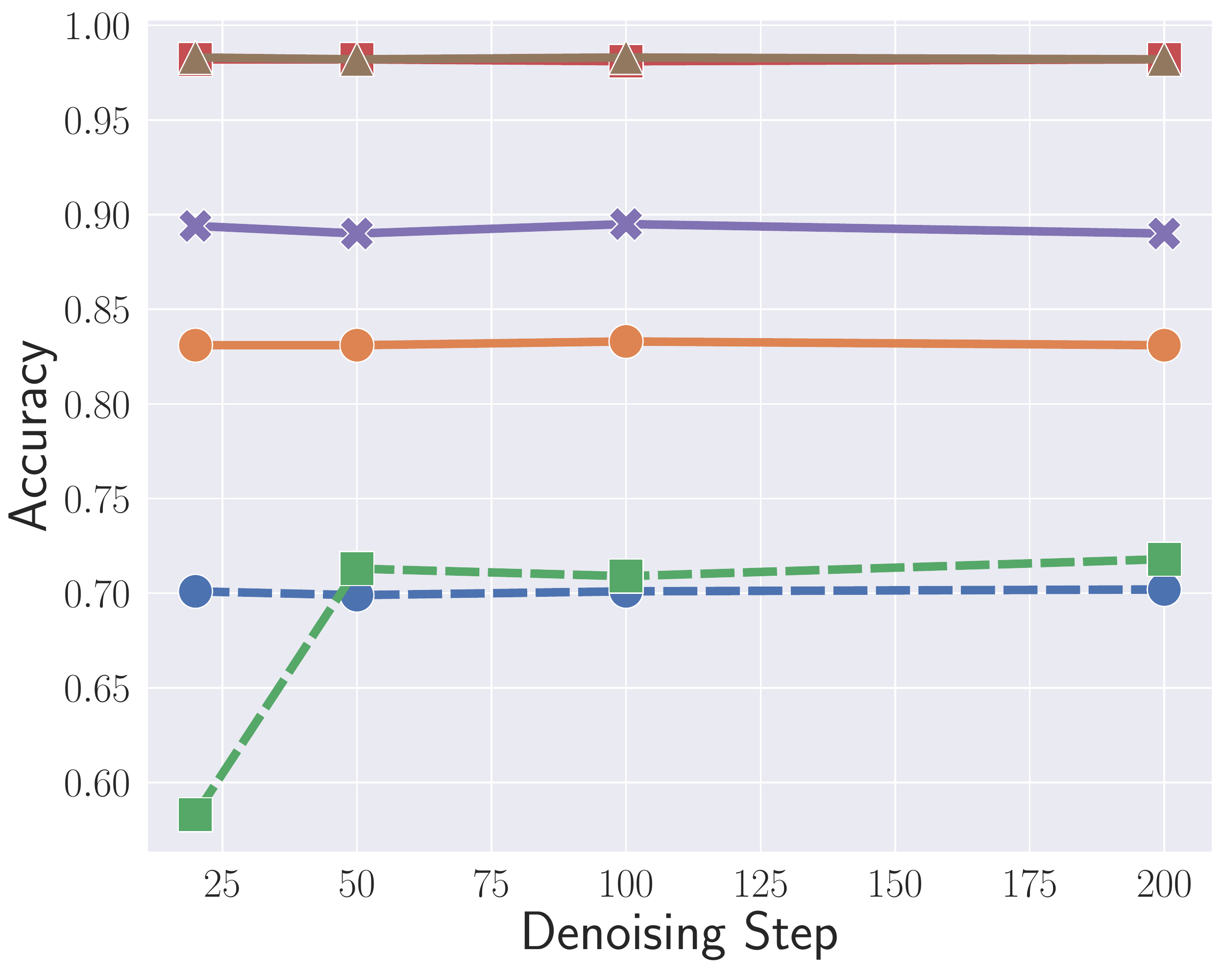}
\caption{VG-Face}
\label{figure:step_vg}
\end{subfigure}
\caption{Test accuracy of the proposed attack methods with varying denoising steps on the LDM model.
The non-member dataset of (a) is MSCOCO-Face and of (b) is VG-Face.
Dashed lines represent pixel-level attacks and solid lines represent semantic-level attacks.}
\label{figure:ddim_step}
\end{figure*}

\begin{figure}[!t]
\centering
\includegraphics[width=0.81\columnwidth]{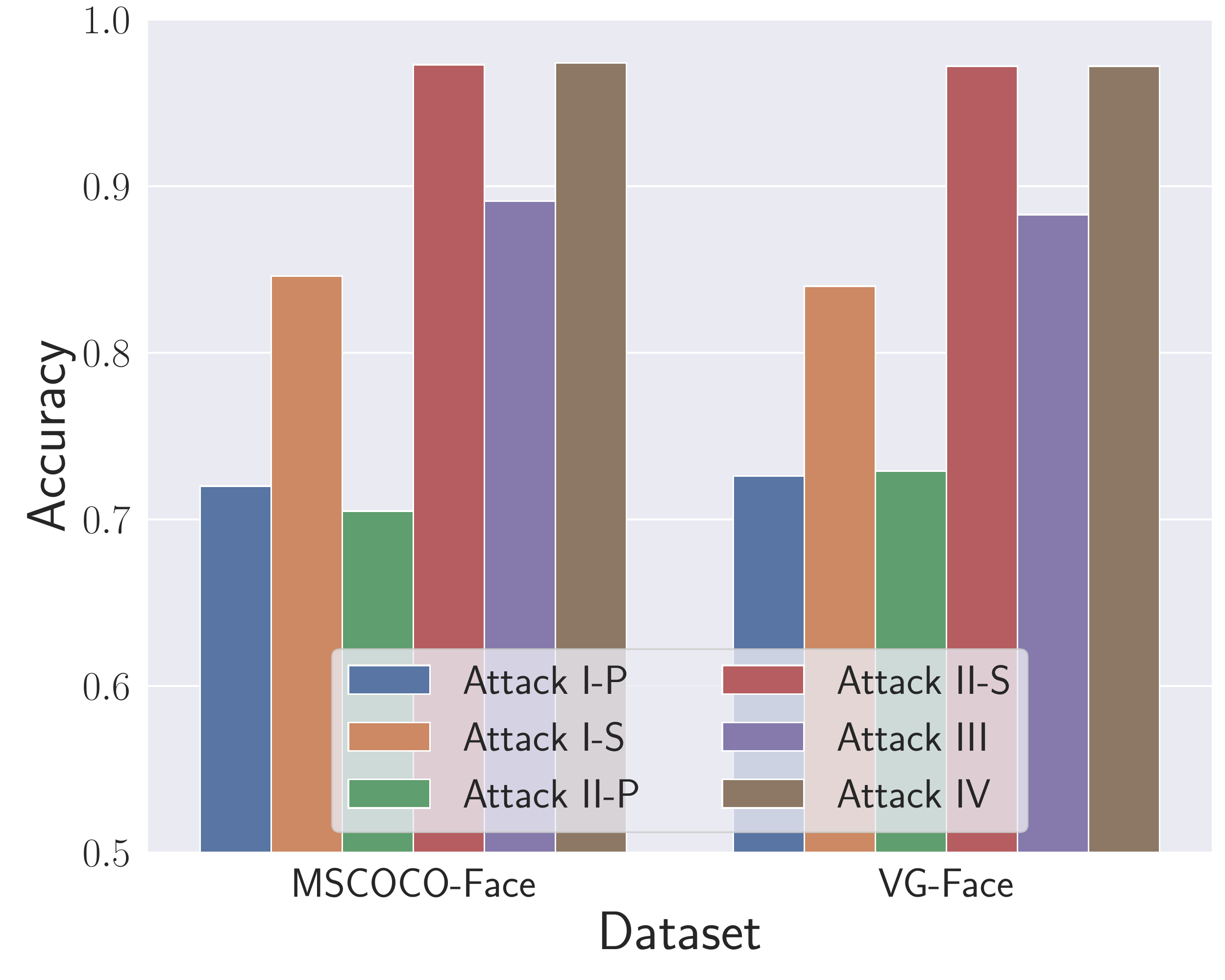}
\caption{Test accuracy of the proposed attack methods on the LDM model.
The caption and embedding generation tools are ClipCap/CLIP.
}
\label{figure:acc_clipcap}
\end{figure}

\mypara{Analysis}
To figure out the reason behind the success of our attacks, we calculate the FID scores between the query images and its corresponding generated images on member and non-member datasets, respectively.
As we leverage BLIP to generate embeddings used for our attacks, we also calculate the FID score based on the BLIP's embeddings.
As illustrated in \autoref{table:fid_score_analysis}, members have a lower FID score than non-members on both cases.
For example, the FID score for Laion-Face is only 9.912, while that of MSCOCO-Face is 20.314.
These results demonstrate that the generated images of members are perceptually closer to the original query images. 
We also plot the histogram of cosine similarity between the same-modality embedding in~\autoref{figure:hist} of~\autoref{section:appendix}, i.e., the query image embedding and generated image embedding, showing the distributions on members and non-members have a clear difference and these same-modality embeddings of the members have a higher cosine similarity in general.

\mypara{Effect of the caption/embedding generation tools}
We investigate if the selection of the caption generation tool and embedding generation tool affects the attack performance.
To this end, we leverage pretrained CLIP to generate embeddings for generated captions, generated images, and query images, and ClipCap~\cite{MHB21}, a state-of-the-art image captioning model, to generate caption for each query image.
As illustrated in~\autoref{figure:acc_clipcap}, the attack performance is consistent with when we use BLIP as the caption and embedding generation tool.
Concretely, Attack IV and Attack II-S still achieve the best performance, with semantic-level attacks still performing better than pixel-level attacks.
Hence, we conclude that the choice of embedding and caption generation tools has negligible effect on the attack performance.

\begin{figure}[!t]
\centering
\includegraphics[width=0.81\columnwidth]{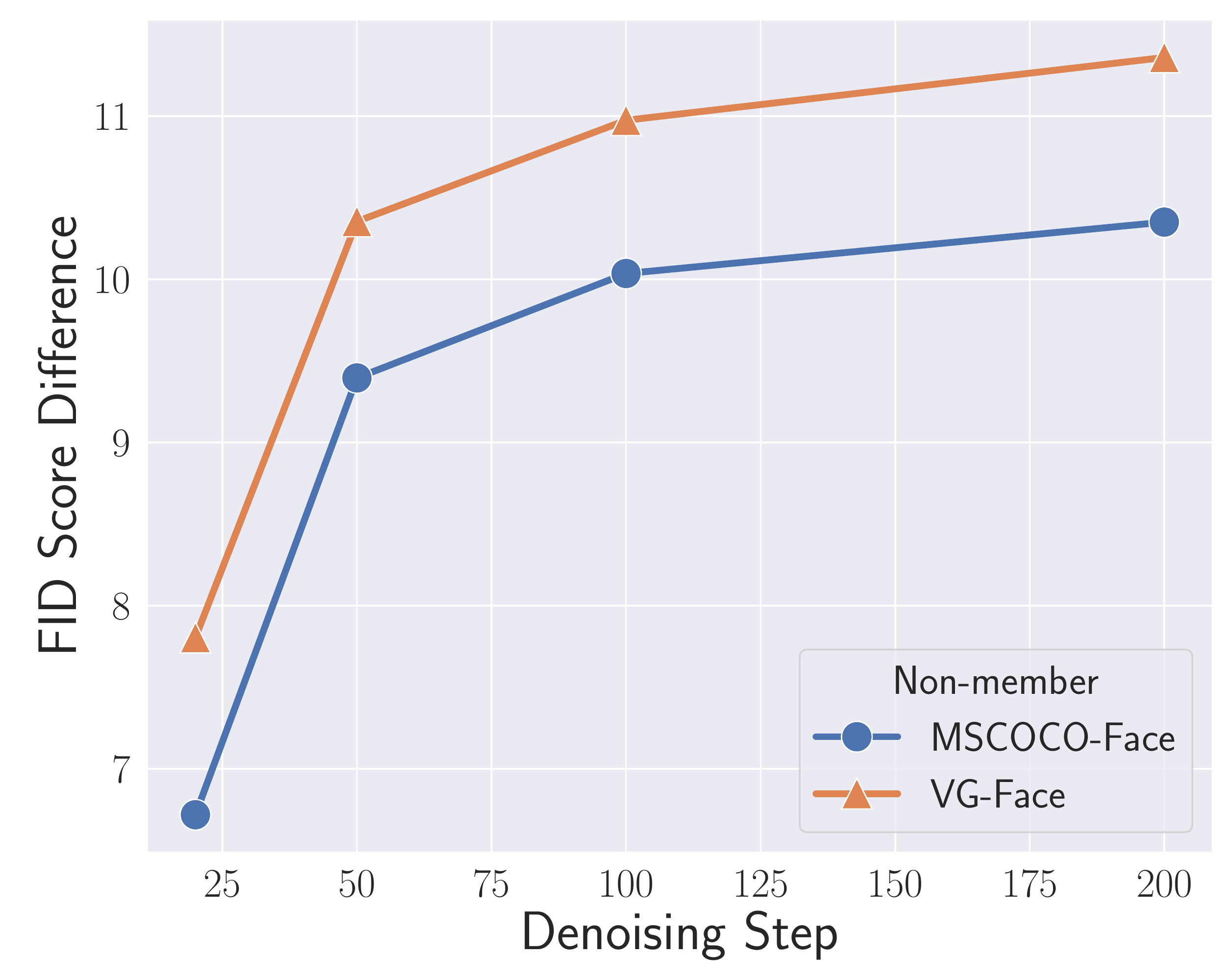}
\caption{FID scores' difference between member and non-member datasets.
}
\label{figure:step_fid}
\end{figure}

\begin{figure*}[!t]
\centering
\begin{subfigure}{0.9\columnwidth}
\includegraphics[width=0.9\columnwidth]{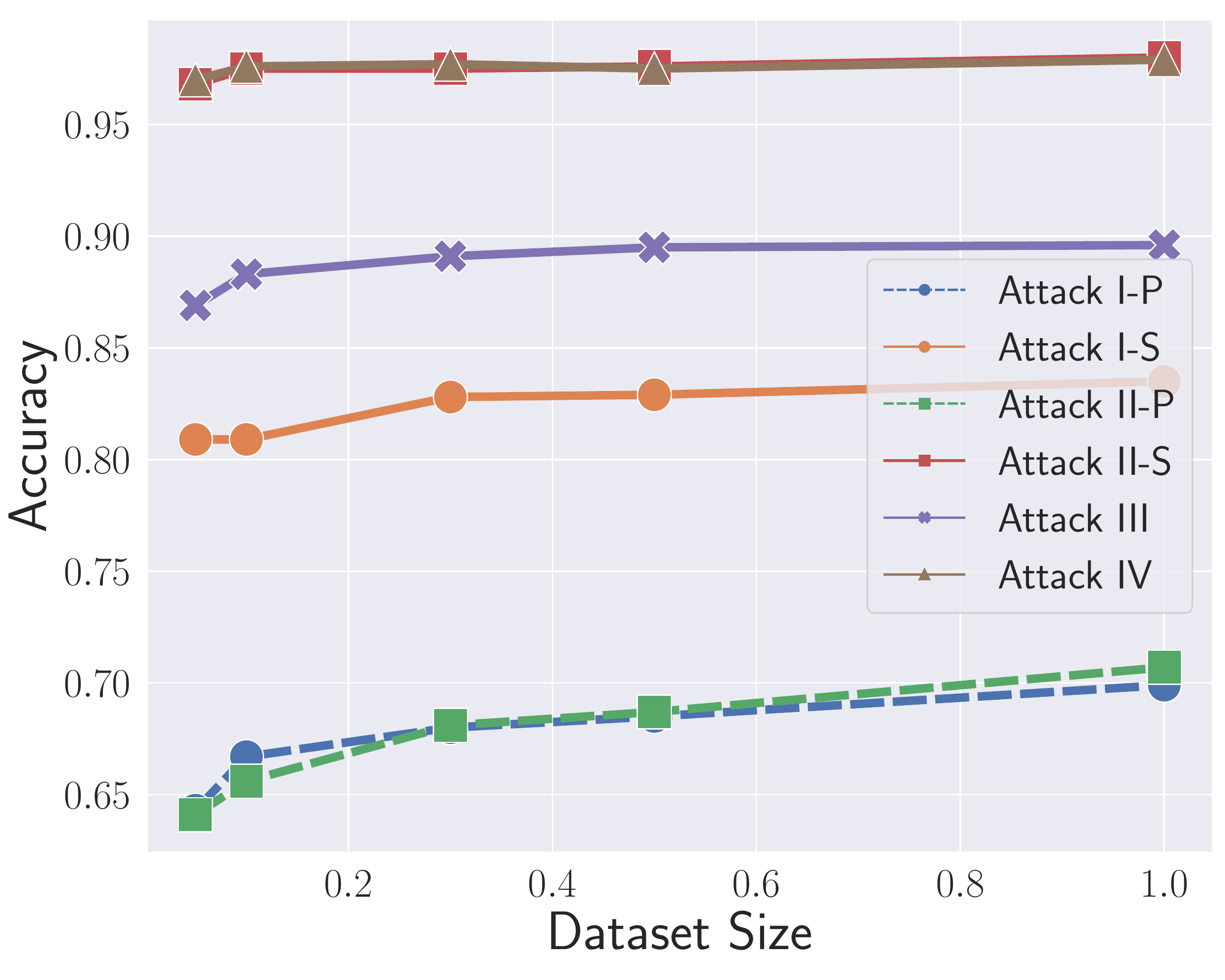}
\caption{MSCOCO-Face}
\label{figure:size_mscoco}
\end{subfigure}
\begin{subfigure}{0.9\columnwidth}
\includegraphics[width=0.9\columnwidth]{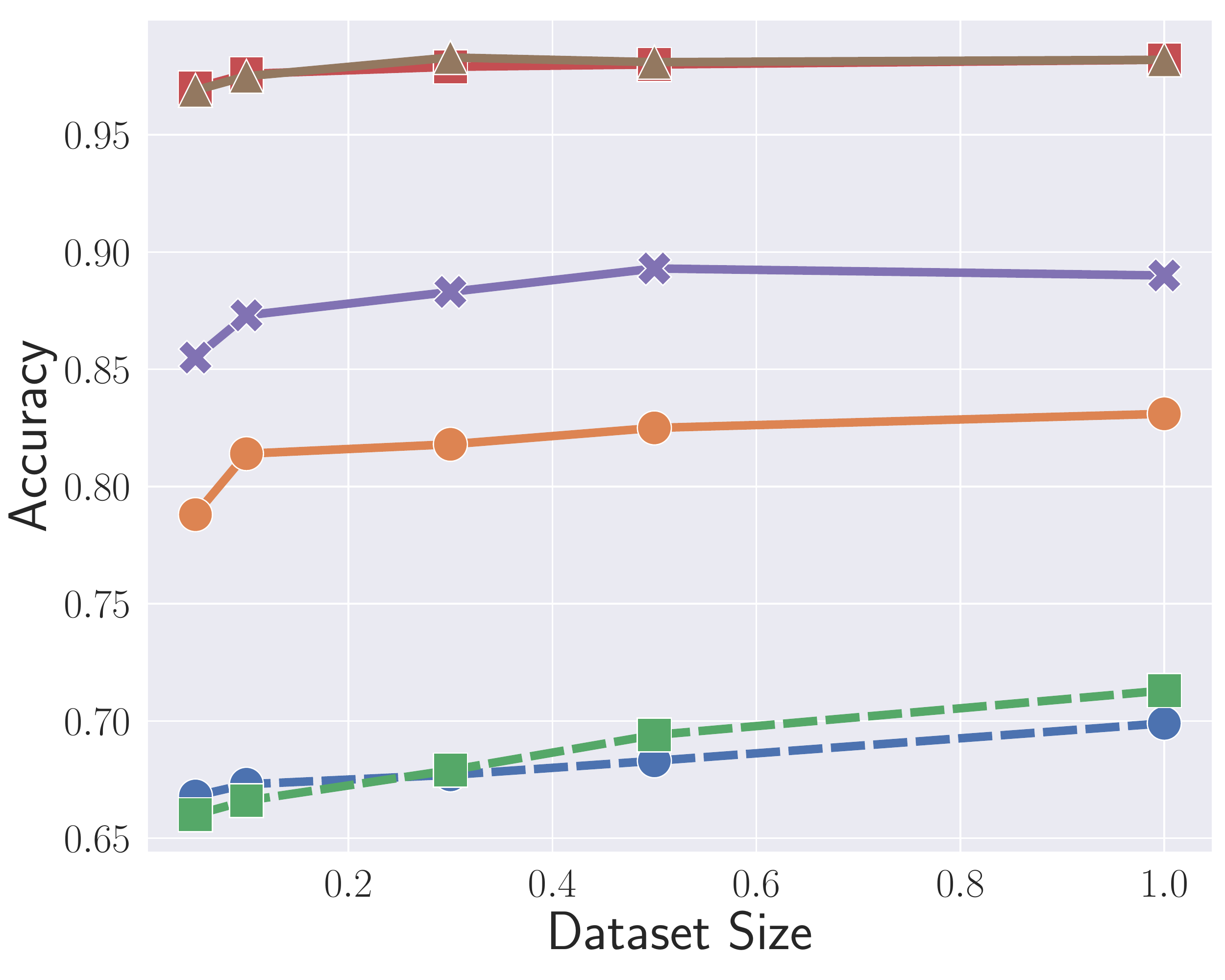}
\caption{VG-Face}
\label{figure:size_vg}
\end{subfigure}
\caption{Test accuracy of the proposed attack methods with varying auxiliary dataset's size on the LDM model.
The non-member dataset of (a) is MSCOCO-Face and of (b) is VG-Face.}
\label{figure:varying_size}
\end{figure*}

\begin{figure*}[!t]
\centering
\begin{subfigure}{0.9\columnwidth}
\includegraphics[width=0.98\columnwidth]{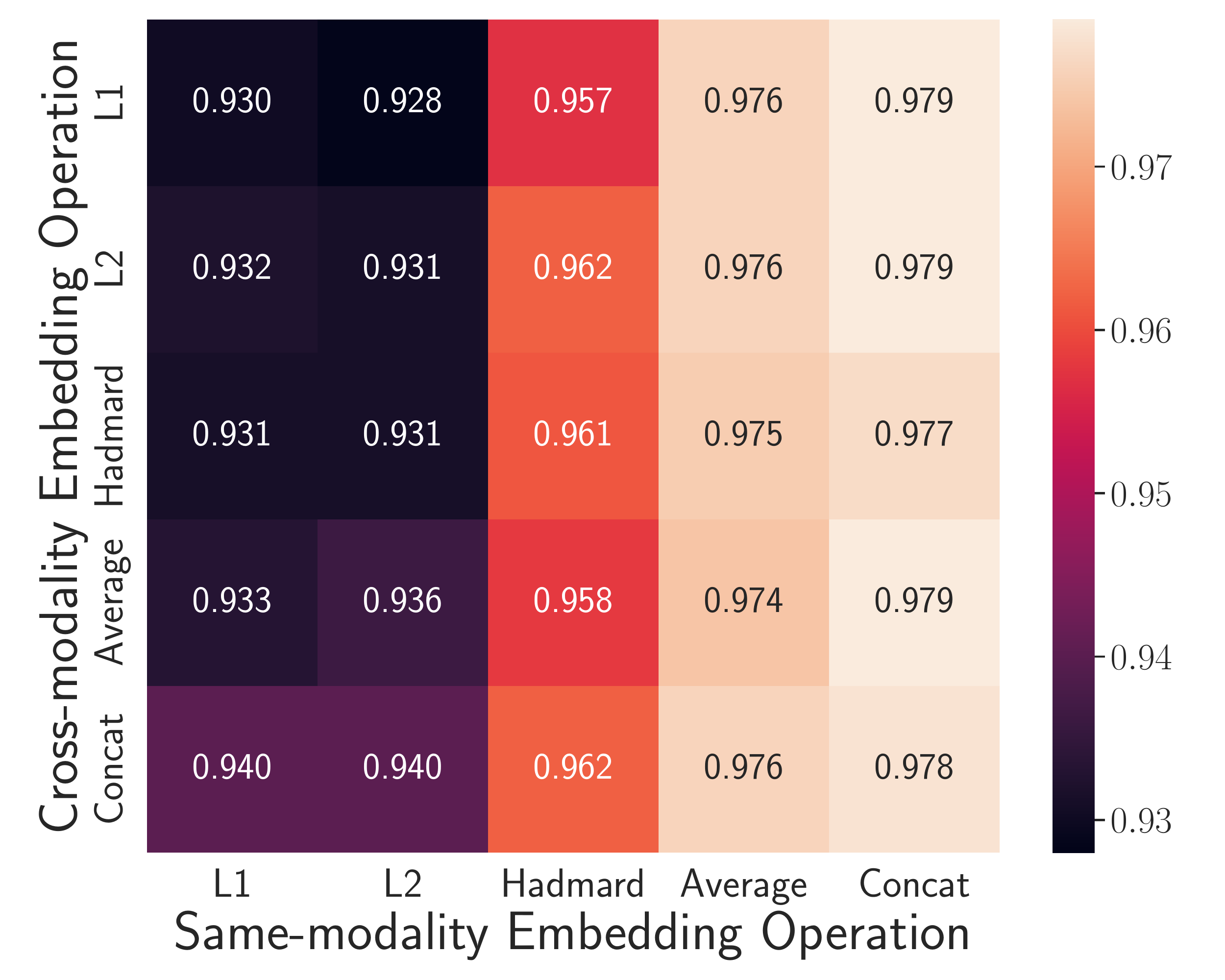}
\caption{MSCOCO-Face}
\label{figure:heatmap_mscoco}
\end{subfigure}
\begin{subfigure}{0.9\columnwidth}
\includegraphics[width=0.98\columnwidth]{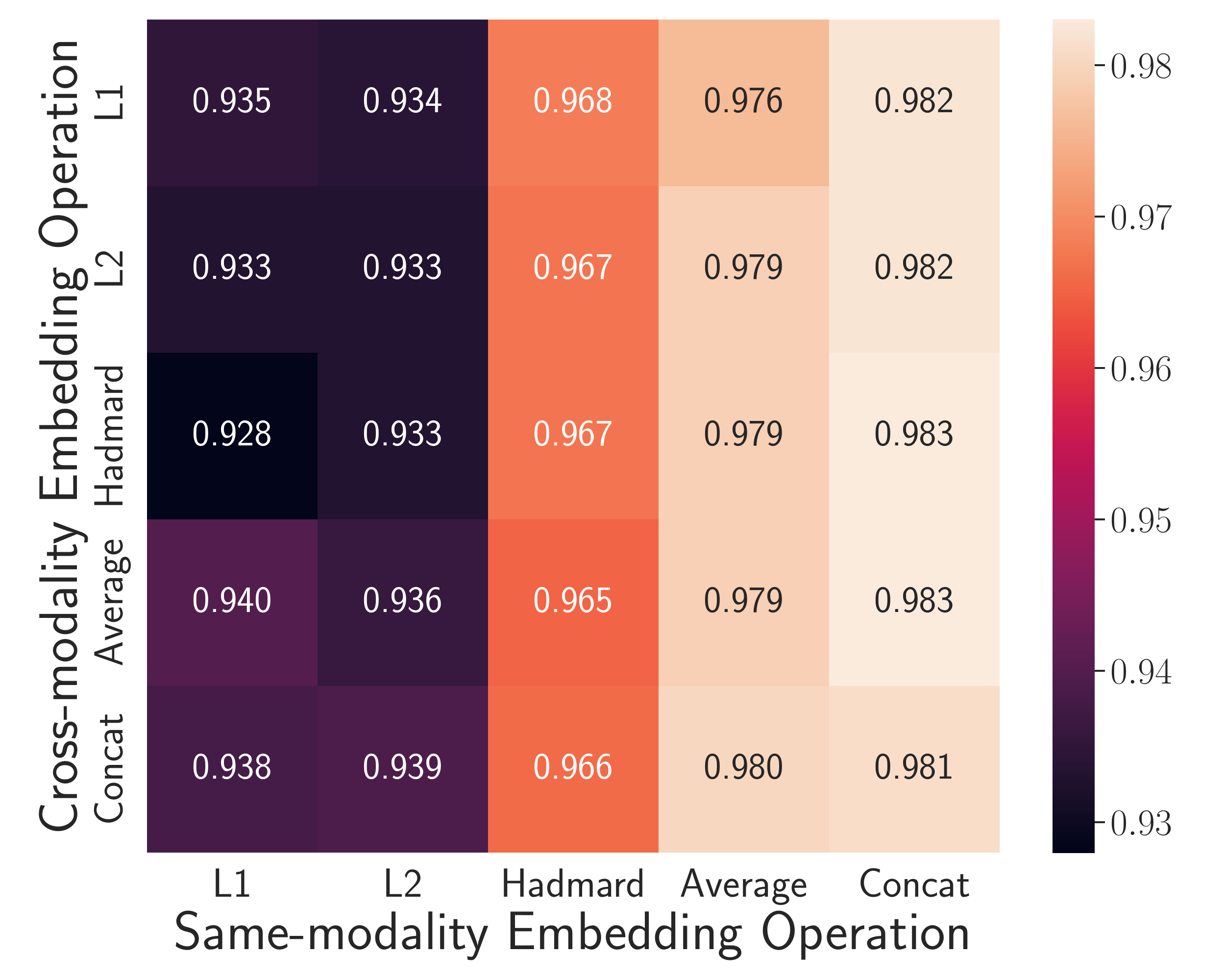}
\caption{VG-Face}
\label{figure:hratmap_vg}
\end{subfigure}
\caption{
Test accuracy of Attack IV on the LDM model with different operations for the cross-modality and same-modality embeddings.
The non-member dataset of (a) is MSCOCO-Face and of (b) is VG-Face.}
\label{figure:heatmap}
\end{figure*}

\mypara{Effect of the denoising steps}
At the inference time of diffusion-based models, we need to manually set the denoising step.
Generally, the more denoising steps, the better the quality of the generated images, which leads to higher time and memory costs.
This motivates us to explore whether more steps can bring better results, or whether a small number of steps is enough to give us good attack performance.
In~\autoref{figure:ddim_step}, we start with 20 denoising steps as the resulting images become visually usable, and gradually increase the steps to 50, 100, and 200, where 50 is the default setting for previous experiments.
We can observe that the proposed attacks work well even with 20 denoising steps except Attack II-P.
Meanwhile, increasing the denoising steps does not gain much improvement of the attack performance.
To explain the results, we calculate the difference between the FID scores of the members and non-members in~\autoref{figure:step_fid}.
We can see that the difference is already very large at 20 denoising steps on both cases.
Although the difference continues increasing with the growth of denoising steps, it is sufficient to enable most of attacks successfully when the generated images are just visually available (e.g., 20 denoising steps in our cases).

\mypara{Effect of the auxiliary dataset's size}
So far, we assume that the adversary can have a small subset $\dtm$ from the training data to construct the auxiliary dataset.
In the real world, it is more likely that the adversary can only get less data.
Therefore, we investigate the influence of auxiliary dataset’ size on attack performance.
Specifically, we randomly sample from the above member $\dtm$ with different proportions $\{0.05, 0.1, 0,3, 0.5, 1\}$ and sample the same size from the local non-member $\dlnm$.
As shown in~\autoref{figure:varying_size}, the proposed attacks are more effective than random guessing even with 5\% auxiliary dataset.
Especially the Attack II-S and Attack IV, the size of auxiliary dataset only has negligible influence on them.
More results on the DALL-E mini model are in~\autoref{figure:dalle_varying_size} of~\autoref{section:appendix}, and the same conclusions can be drawn.

\mypara{Operations for processing embeddings}
In Attack II-S/III/IV, we calculate the element-wise distance between two embeddings.
Besides leveraging traditional metrics. e.g., L1 distance and L2 distance, we also consider other three pair-wise operations.
These operations are summarized in~\autoref{table:pairwise} of~\autoref{section:appendix}.
We refer to the embedding distance between the generated images and query images as same-modality embedding and to the embedding distance between the generated images and generated captions as cross-modality embedding.
We explore the effect of different operations on Attack IV, as it takes both the same-modality and cross-modality embedding as input.
As illustrated in~\autoref{figure:heatmap}, the operation of the same-modality embedding plays a decisive role. 
The concatenation operation for the same-modality embedding brings the best performance of Attack IV, as it achieves over 97.7\% test accuracy on both cases no matter the operation for the cross-modality embedding.
We observe the concatenation operation is also the most appropriate one for the cross-modality embedding, as it can increase the attack performance when the same-modality embedding is less dominant.
Thus, we use the concatenation operation for Attack II-S, Attack III, and Attack IV.

\section{Defense}

To mitigate membership leakage, the traditional way is to only output top-k predictions or the predicted labels. 
But this method cannot be extended to our scenarios as the output format of the text-to-image generation model is image.
Chen et al.\ apply differential private (DP) to the target GAN model as a defense mechanism~\cite{CYZF20}.
However, due to the resource limitation, i.e., insufficient GPU resource, we are unable to train a text-to-image generation model with DP from scratch.
In the previous experiment, we limit the number of data samples from the target training data.
We can observe that by reducing the size of the dataset to 5\% of its original size, the proposed attacks only have a slight deterioration and are still effective. 
Hence, we cannot limit the attack performance via restricting the number of member samples the adversary can have.
Our future work will concentrate on thoroughly exploring more effective defense mechanisms against our attacks.

\section{Conclusion}

We take the first step towards studying membership leakage in text-to-image generation models, where an adversary aims to infer whether a given image is used to train a target text-to-image generation model.
Based on the characteristics of the text-to-image generation models, we consider three key intuitions and design four attack methods accordingly.
We conduct comprehensive experiments on two representative text-to-image generation modes.
Empirical results show that all proposed attacks achieve remarkable performance, which convincingly demonstrates that membership leakage is a severe threat to the text-to-image generation models.
Furthermore, to deeply investigate which factors and their effects on the attack performance, we conduct an in-depth ablation study from different perspectives that can guide developers and researchers to be alert to vulnerabilities in text-to-image generation models.

\bibliography{normal_generated_py3.bib}

\begin{thebibliography}{10}

\bibitem{ABCPK18}
Yossi Adi, Carsten Baum, Moustapha Cisse, Benny Pinkas, and Joseph Keshet.
\newblock {Turning Your Weakness Into a Strength: Watermarking Deep Neural
  Networks by Backdooring}.
\newblock In {\em {USENIX Security Symposium (USENIX Security)}}, pages
  1615--1631. USENIX, 2018.

\bibitem{CSDS21}
Soravit Changpinyo, Piyush Sharma, Nan Ding, and Radu Soricut.
\newblock {Conceptual 12M: Pushing Web-Scale Image-Text Pre-Training To
  Recognize Long-Tail Visual Concepts}.
\newblock In {\em {IEEE Conference on Computer Vision and Pattern Recognition
  (CVPR)}}, pages 3558--3568. IEEE, 2021.

\bibitem{CYZF20}
Dingfan Chen, Ning Yu, Yang Zhang, and Mario Fritz.
\newblock {GAN-Leaks: A Taxonomy of Membership Inference Attacks against
  Generative Models}.
\newblock In {\em {ACM SIGSAC Conference on Computer and Communications
  Security (CCS)}}, pages 343--362. ACM, 2020.

\bibitem{CFLVGDZ15}
Xinlei Chen, Hao Fang, Tsung{-}Yi Lin, Ramakrishna Vedantam, Saurabh Gupta,
  Piotr Doll{\'{a}}r, and C.~Lawrence Zitnick.
\newblock {Microsoft COCO Captions: Data Collection and Evaluation Server}.
\newblock {\em {CoRR abs/1504.00325}}, 2015.

\bibitem{CTCP21}
Christopher A.~Choquette Choo, Florian Tram{\`e}r, Nicholas Carlini, and
  Nicolas Papernot.
\newblock {Label-Only Membership Inference Attacks}.
\newblock In {\em {International Conference on Machine Learning (ICML)}}, pages
  1964--1974. PMLR, 2021.

\bibitem{DGZYKZ20}
Jiankang Deng, Jia Guo, Yuxiang Zhou, Jinke Yu, Irene Kotsia, and Stefanos
  Zafeiriou.
\newblock {RetinaFace: Single-stage Dense Face Localisation in the Wild}.
\newblock In {\em {IEEE Conference on Computer Vision and Pattern Recognition
  (CVPR)}}, pages 5203--5212. IEEE, 2020.

\bibitem{DYHZZYLZSYT21}
Ming Ding, Zhuoyi Yang, Wenyi Hong, Wendi Zheng, Chang Zhou, Da~Yin, Junyang
  Lin, Xu~Zou, Zhou Shao, Hongxia Yang, and Jie Tang.
\newblock {CogView: Mastering Text-to-Image Generation via Transformers}.
\newblock In {\em {Annual Conference on Neural Information Processing Systems
  (NeurIPS)}}, pages 19822--19835. NeurIPS, 2021.

\bibitem{HLXCZ22}
Xinlei He, Zheng Li, Weilin Xu, Cory Cornelius, and Yang Zhang.
\newblock {Membership-Doctor: Comprehensive Assessment of Membership Inference
  Against Machine Learning Models}.
\newblock {\em {CoRR abs/2208.10445}}, 2022.

\bibitem{HWWBSZ21}
Xinlei He, Rui Wen, Yixin Wu, Michael Backes, Yun Shen, and Yang Zhang.
\newblock {Node-Level Membership Inference Attacks Against Graph Neural
  Networks}.
\newblock {\em {CoRR abs/2102.05429}}, 2021.

\bibitem{HHB19}
Benjamin Hilprecht, Martin H{\"{a}}rterich, and Daniel Bernau.
\newblock {Monte Carlo and Reconstruction Membership Inference Attacks against
  Generative Models}.
\newblock {\em {Privacy Enhancing Technologies Symposium}}, 2019.

\bibitem{HYYBGC21}
Bo~Hui, Yuchen Yang, Haolin Yuan, Philippe Burlina, Neil~Zhenqiang Gong, and
  Yinzhi Cao.
\newblock {Practical Blind Membership Inference Attack via Differential
  Comparisons}.
\newblock In {\em {Network and Distributed System Security Symposium (NDSS)}}.
  Internet Society, 2021.

\bibitem{KZGJHKCKLSBF17}
Ranjay Krishna, Yuke Zhu, Oliver Groth, Justin Johnson, Kenji Hata, Joshua
  Kravitz, Stephanie Chen, Yannis Kalantidis, Li{-}Jia Li, David~A. Shamma,
  Michael~S. Bernstein, and Li~Fei{-}Fei.
\newblock {Visual Genome: Connecting Language and Vision Using Crowdsourced
  Dense Image Annotations}.
\newblock {\em {International Journal of Computer Vision}}, 2017.

\bibitem{LLXH22}
Junnan Li, Dongxu Li, Caiming Xiong, and Steven C.~H. Hoi.
\newblock {{BLIP:} Bootstrapping Language-Image Pre-training for Unified
  Vision-Language Understanding and Generation}.
\newblock {\em {CoRR abs/2201.12086}}, 2022.

\bibitem{LHZG19}
Zheng Li, Chengyu Hu, Yang Zhang, and Shanqing Guo.
\newblock {How to Prove Your Model Belongs to You: A Blind-Watermark based
  Framework to Protect Intellectual Property of DNN}.
\newblock In {\em {Annual Computer Security Applications Conference (ACSAC)}},
  pages 126--137. ACM, 2019.

\bibitem{LLHYBZ222}
Zheng Li, Yiyong Liu, Xinlei He, Ning Yu, Michael Backes, and Yang Zhang.
\newblock {Auditing Membership Leakages of Multi-Exit Networks}.
\newblock {\em {CoRR abs/2208.11180}}, 2022.

\bibitem{LZ21}
Zheng Li and Yang Zhang.
\newblock {Membership Leakage in Label-Only Exposures}.
\newblock In {\em {ACM SIGSAC Conference on Computer and Communications
  Security (CCS)}}, pages 880--895. ACM, 2021.

\bibitem{LZBZ22}
Yiyong Liu, Zhengyu Zhao, Michael Backes, and Yang Zhang.
\newblock {Membership Inference Attacks by Exploiting Loss Trajectory}.
\newblock {\em {CoRR abs/2208.14933}}, 2022.

\bibitem{LWHSZBCFZ22}
Yugeng Liu, Rui Wen, Xinlei He, Ahmed Salem, Zhikun Zhang, Michael Backes,
  Emiliano~De Cristofaro, Mario Fritz, and Yang Zhang.
\newblock {ML-Doctor: Holistic Risk Assessment of Inference Attacks Against
  Machine Learning Models}.
\newblock In {\em {USENIX Security Symposium (USENIX Security)}}. USENIX, 2022.

\bibitem{MHB21}
Ron Mokady, Amir Hertz, and Amit~H. Bermano.
\newblock {ClipCap: {CLIP} Prefix for Image Captioning}.
\newblock {\em {CoRR abs/2111.09734}}, 2021.

\bibitem{NDRSMMSC21}
Alex Nichol, Prafulla Dhariwal, Aditya Ramesh, Pranav Shyam, Pamela Mishkin,
  Bob McGrew, Ilya Sutskever, and Mark Chen.
\newblock {GLIDE: Towards Photorealistic Image Generation and Editing with
  Text-Guided Diffusion Models}.
\newblock {\em {CoRR abs/2112.10741}}, 2021.

\bibitem{ONK21}
Iyiola~E. Olatunji, Wolfgang Nejdl, and Megha Khosla.
\newblock {Membership Inference Attack on Graph Neural Networks}.
\newblock {\em {CoRR abs/2101.06570}}, 2021.

\bibitem{RKHRGASAMCKS21}
Alec Radford, Jong~Wook Kim, Chris Hallacy, Aditya Ramesh, Gabriel Goh,
  Sandhini Agarwal, Girish Sastry, Amanda Askell, Pamela Mishkin, Jack Clark,
  Gretchen Krueger, and Ilya Sutskever.
\newblock {Learning Transferable Visual Models From Natural Language
  Supervision}.
\newblock In {\em {International Conference on Machine Learning (ICML)}}, pages
  8748--8763. PMLR, 2021.

\bibitem{RDNCC22}
Aditya Ramesh, Prafulla Dhariwal, Alex Nichol, Casey Chu, and Mark Chen.
\newblock {Hierarchical Text-Conditional Image Generation with CLIP Latents}.
\newblock {\em {CoRR abs/2204.06125}}, 2022.

\bibitem{RPGGVRCS21}
Aditya Ramesh, Mikhail Pavlov, Gabriel Goh, Scott Gray, Chelsea Voss, Alec
  Radford, Mark Chen, and Ilya Sutskever.
\newblock {Zero-Shot Text-to-Image Generation}.
\newblock In {\em {International Conference on Machine Learning (ICML)}}, pages
  8821--8831. JMLR, 2021.

\bibitem{R17}
Jason~Tyler Rolfe.
\newblock {Discrete Variational Autoencoders}.
\newblock In {\em {International Conference on Learning Representations
  (ICLR)}}, 2017.

\bibitem{RBLEO22}
Robin Rombach, Andreas Blattmann, Dominik Lorenz, Patrick Esser, and
  Bj{\"{o}}rn Ommer.
\newblock {High-Resolution Image Synthesis with Latent Diffusion Models}.
\newblock In {\em {IEEE Conference on Computer Vision and Pattern Recognition
  (CVPR)}}, pages 10684--10695. IEEE, 2022.

\bibitem{RCK18}
Bita~Darvish Rouhani, Huili Chen, and Farinaz Koushanfar.
\newblock {DeepSigns: A Generic Watermarking Framework for IP Protection of
  Deep Learning Models}.
\newblock {\em {CoRR abs/1804.00750}}, 2018.

\bibitem{SCSLWDGAMLSHFN22}
Chitwan Saharia, William Chan, Saurabh Saxena, Lala Li, Jay Whang, Emily
  Denton, Seyed Kamyar~Seyed Ghasemipour, Burcu~Karagol Ayan, S.~Sara Mahdavi,
  Rapha~Gontijo Lopes, Tim Salimans, Jonathan Ho, David~J. Fleet, and Mohammad
  Norouzi.
\newblock {Photorealistic Text-to-Image Diffusion Models with Deep Language
  Understanding}.
\newblock {\em {CoRR abs/2205.11487}}, 2022.

\bibitem{SZHBFB19}
Ahmed Salem, Yang Zhang, Mathias Humbert, Pascal Berrang, Mario Fritz, and
  Michael Backes.
\newblock {ML-Leaks: Model and Data Independent Membership Inference Attacks
  and Defenses on Machine Learning Models}.
\newblock In {\em {Network and Distributed System Security Symposium (NDSS)}}.
  Internet Society, 2019.

\bibitem{SVBKMKCJK21}
Christoph Schuhmann, Richard Vencu, Romain Beaumont, Robert Kaczmarczyk,
  Clayton Mullis, Aarush Katta, Theo Coombes, Jenia Jitsev, and Aran
  Komatsuzaki.
\newblock {{LAION-400M:} Open Dataset of CLIP-Filtered 400 Million Image-Text
  Pairs}.
\newblock {\em {CoRR abs/2111.02114}}, 2021.

\bibitem{SDGS18}
Piyush Sharma, Nan Ding, Sebastian Goodman, and Radu Soricut.
\newblock {Conceptual Captions: {A} Cleaned, Hypernymed, Image Alt-text Dataset
  For Automatic Image Captioning}.
\newblock In {\em {Annual Meeting of the Association for Computational
  Linguistics (ACL)}}, pages 2556--2565. ACL, 2018.

\bibitem{SSSS17}
Reza Shokri, Marco Stronati, Congzheng Song, and Vitaly Shmatikov.
\newblock {Membership Inference Attacks Against Machine Learning Models}.
\newblock In {\em {IEEE Symposium on Security and Privacy (S\&P)}}, pages
  3--18. IEEE, 2017.

\bibitem{TSFENPBL16}
Bart Thomee, David~A. Shamma, Gerald Friedland, Benjamin Elizalde, Karl Ni,
  Douglas Poland, Damian Borth, and Li{-}Jia Li.
\newblock {{YFCC100M:} the new data in multimedia research}.
\newblock {\em {Commun. {ACM}}}, 2016.

\bibitem{UNSS17}
Yusuke Uchida, Yuki Nagai, Shigeyuki Sakazawa, and Shin'ichi Satoh.
\newblock {Embedding Watermarks into Deep Neural Networks}.
\newblock In {\em {International Conference on Multimedia Retrieval (ICMR)}},
  pages 269--277. ACM, 2017.

\bibitem{WYPY21}
Bang Wu, Xiangwen Yang, Shirui Pan, and Xingliang Yuan.
\newblock {Adapting Membership Inference Attacks to {GNN} for Graph
  Classification: Approaches and Implications}.
\newblock In {\em {International Conference on Data Mining (ICDM)}}. IEEE,
  2021.

\bibitem{YXKLBWVKYAHHPLZBW22}
Jiahui Yu, Yuanzhong Xu, Jing~Yu Koh, Thang Luong, Gunjan Baid, Zirui Wang,
  Vijay Vasudevan, Alexander Ku, Yinfei Yang, Burcu~Karagol Ayan, Ben
  Hutchinson, Wei Han, Zarana Parekh, Xin Li, Han Zhang, Jason Baldridge, and
  Yonghui Wu.
\newblock {Scaling Autoregressive Models for Content-Rich Text-to-Image
  Generation}.
\newblock {\em {CoRR abs/2206.10789}}, 2022.

\bibitem{ZGJWSHM18}
Jialong Zhang, Zhongshu Gu, Jiyong Jang, Hui Wu, Marc~Ph. Stoecklin, Heqing
  Huang, and Ian Molloy.
\newblock {Protecting Intellectual Property of Deep Neural Networks with
  Watermarking}.
\newblock In {\em {ACM Asia Conference on Computer and Communications Security
  (ASIACCS)}}, pages 159--172. ACM, 2018.

\bibitem{ZYZBCHYCZW22}
Yinglin Zheng, Hao Yang, Ting Zhang, Jianmin Bao, Dongdong Chen, Yangyu Huang,
  Lu~Yuan, Dong Chen, Ming Zeng, and Fang Wen.
\newblock {General Facial Representation Learning in a Visual-Linguistic
  Manner}.
\newblock In {\em {IEEE Conference on Computer Vision and Pattern Recognition
  (CVPR)}}, pages 18697--18709. IEEE, 2022.

\end{thebibliography}
\bibliographystyle{plain}

\newpage

\appendix
\section{Appendix}
\label{section:appendix}

\mypara{Member and non-member datasets creation for LDM}
We first randomly sample 30K images from the LAION-Face dataset as the member dataset.
To be in line with the member dataset, we generate the MSCOCO-Face and VG-Face datasets by following the generation process of the Laion-Face, i.e., adopting RetinaFace~\cite{DGZYKZ20} as a face detector to detect whether a given image contains a human face and filtering out those whose face detection scores are less than 0.9.
To maintain the data balance, we randomly sample 30K image-text pairs from MSCOCO-Face as the final version of the non-member dataset.
As the VG-Face only contains 26,134 image-text pairs, we resize the member dataset to the equal size when considering VG-Face as the non-member dataset.
Note that we only leverage images from these image-text-pair datasets, and we can also launch these attacks by using image-only datasets.

\mypara{Member and non-member datasets creation for DALL-E mini}
As it is claimed that faces in general are not generated properly by the DALL-E mini model, we create a 30K subset for each dataset by leveraging the RetinaFace detector again and randomly sampling from images that no face is detected.
The member dataset is referred to as CC3M-No-Face, and the non-member datasets are referred to as MSCOCO-No-Face and VG-No-Face, respectively.

\begin{figure*}[!t]
\centering
\begin{subfigure}{0.9\columnwidth}
\includegraphics[width=0.9\columnwidth]{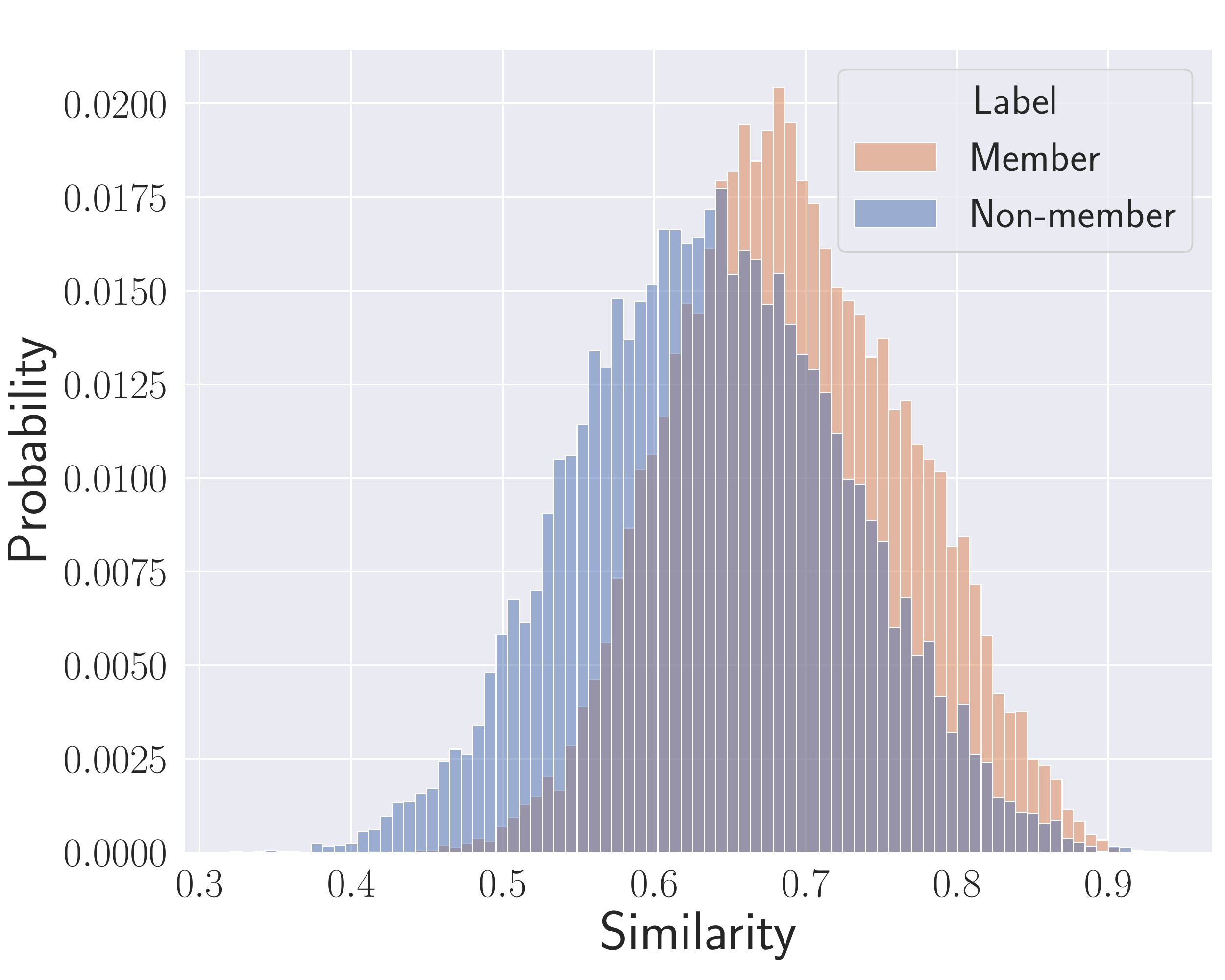}
\caption{MSCOCO-Face}
\label{figure:hist_mscoco}
\end{subfigure}
\begin{subfigure}{0.9\columnwidth}
\includegraphics[width=0.9\columnwidth]{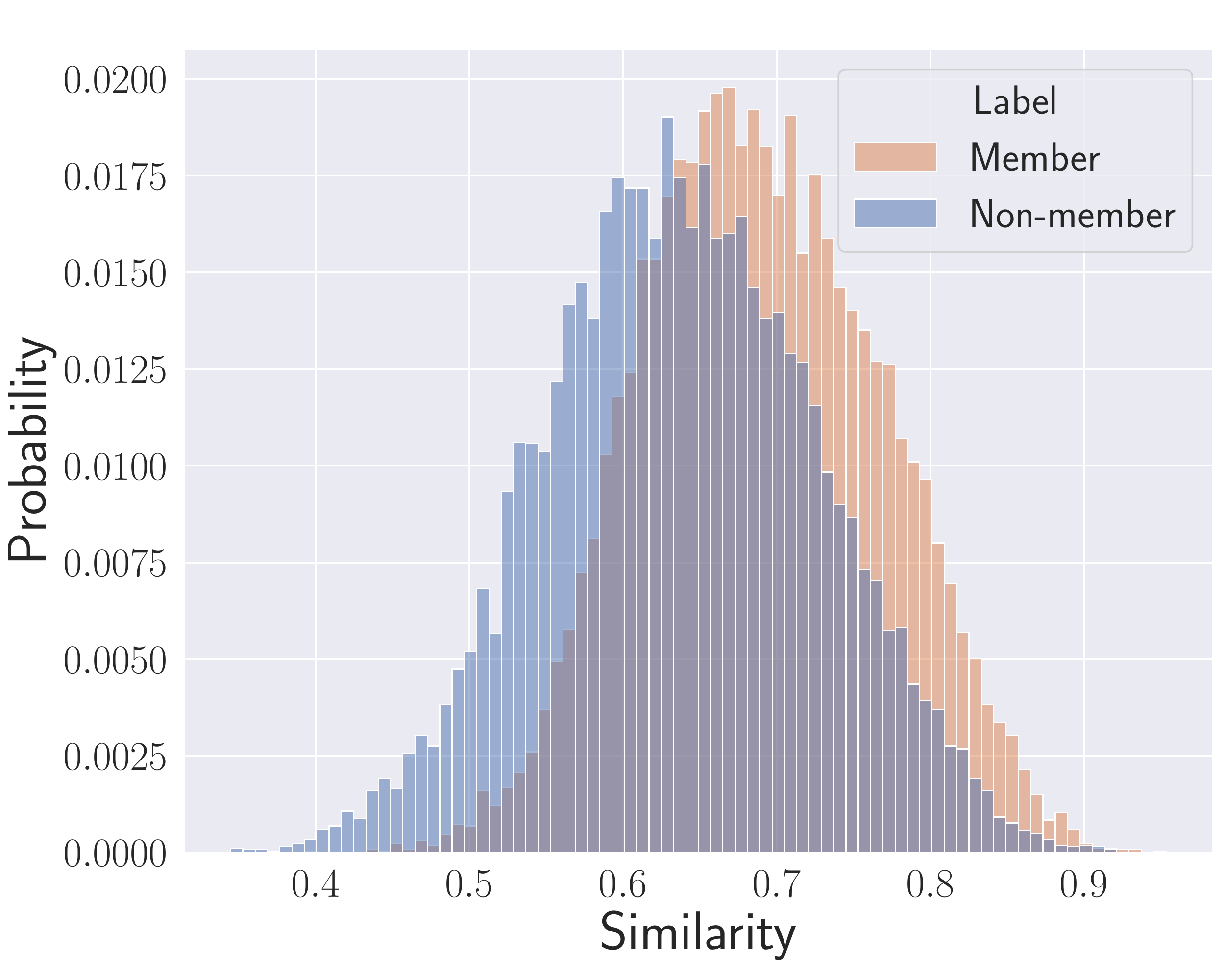}
\caption{VG-Face}
\label{figure:hist_vg}
\end{subfigure}
\caption{Histogram of the cosine similarity between the query image embeddings and generated image embeddings obtained by BLIP.
The text-to-image generation model is fix to LDM.}
\label{figure:hist}
\end{figure*}

\mypara{The cosine similarity between same-modality embedding}
We plot the histogram of the cosine similarity between the query image embeddings and generated image embeddings obtained by BLIP.
As shown in~\autoref{figure:hist}, the distributions of cosine similarity between same-modality embeddings on member dataset and non-member dataset has a clear difference on these two cases.
In addition, these two embeddings of the member dataset have a higher cosine similarity in general, which again verifies the intuition that the distance between generated image $\ig$ and query image $\iq$ from the member dataset should be closer than that from non-member dataset.

\begin{figure*}[!t]
\centering
\begin{subfigure}{0.9\columnwidth}
\includegraphics[width=0.9\columnwidth]{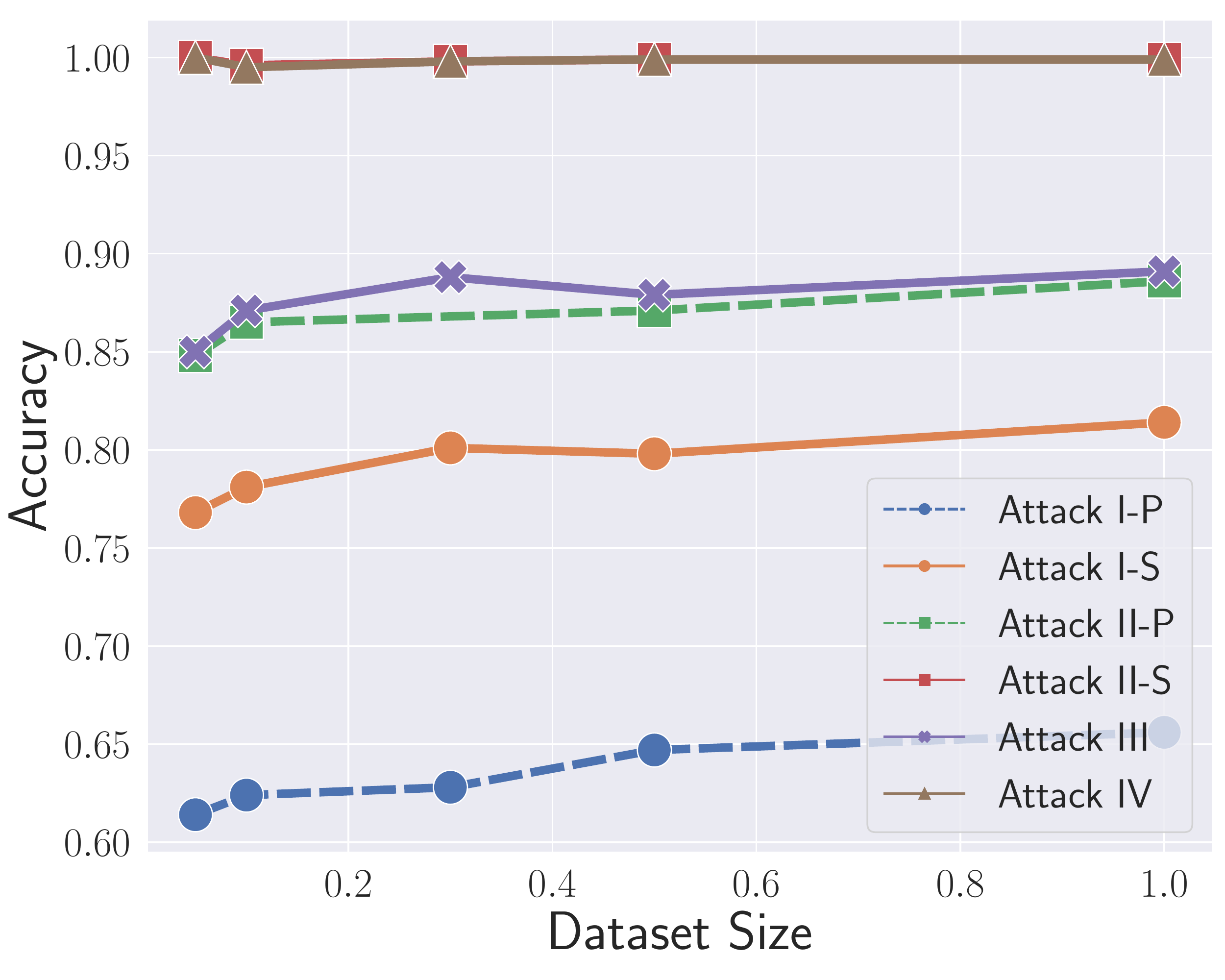}
\caption{MSCOCO-Face}
\label{figure:dalle_size_mscoco}
\end{subfigure}
\begin{subfigure}{0.9\columnwidth}
\includegraphics[width=0.9\columnwidth]{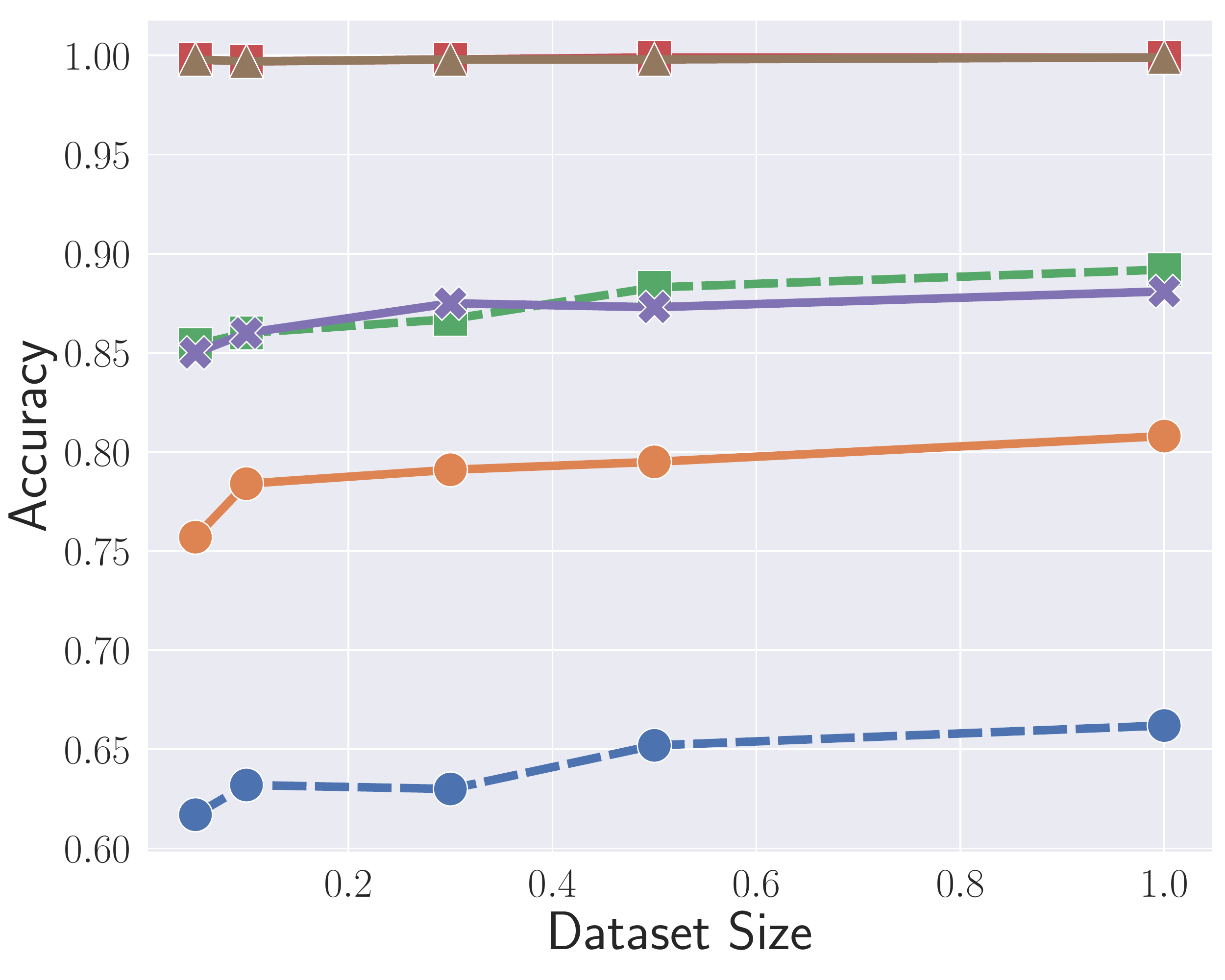}
\caption{VG-Face}
\label{figure:dalle_size_vg}
\end{subfigure}
\caption{Test accuracy of the proposed attack methods with varying auxiliary dataset's size on the DALL-E mini model.
The non-member dataset of (a) is MSCOCO-Face and of (b) is VG-Face.}
\label{figure:dalle_varying_size}
\end{figure*}

\begin{table}[ht]
\caption{Operations}
\label{table:pairwise}
\centering
\renewcommand{\arraystretch}{1.2}
\begin{tabular}{l | c }
\toprule
Operator & Definition\\
\midrule
L1 Distance & $|f_{i}(u) - f_{i}(v)|$ \\
L2 Distance & $|f_{i}(u) - f_{i}(v)|^2$\\
Hadamard & $f_{i}(u) * f_{i}(v)$ \\
Average & $\frac{f_{i}(u) + f_{i}(v)}{2}$ \\
Concatenation & $[f_{i}(u), f_{i}(v)]$\\
\bottomrule
\end{tabular}
\end{table}
\end{document}